\documentstyle[editedvolume]{crckapb}

\def\lsim{\lower.5ex\hbox{$\; \buildrel < \over \sim \;$}}
\def\gsim{\lower.5ex\hbox{$\; \buildrel > \over \sim \;$}}
\def\he#1{\hbox{$^{#1}{\rm He}$}}
\def\li#1{\hbox{$^{#1}{\rm Li}$}}
\def\b1#1{\hbox{$^{1#1}{\rm B}$}}
\def\be#1{\hbox{$^{#1}{\rm Be}$}}
\def\c1#1{\hbox{$^{1#1}{\rm C}$}}
\def\n1#1{\hbox{$^{1#1}{\rm N}$}}
\def\o1#1{\hbox{$^{1#1}{\rm O}$}}
\def\etal{{\it et al.}~}
\begin{opening}
\title{COSMIC LITHIUM-BERYLLIUM-BORON STORY}
\author {Elisabeth Vangioni-Flam}
\institute{ Institut d'Astrophysique de Paris, CNRS\\
98 bis bd Arago Paris France}
\author {Michel Cass\'e}
\institute{  Service d'Astrophysique, CEA\\
Orme des Merisiers\\
91191 Gif/Yvette France\\
 and Institut d'Astrophysique de Paris}
\end{opening}
\runningtitle{COSMIC LITHIUM-BERYLLIUM-BORE STORY}

\begin{document}

\begin{abstract}

Light element nucleosynthesis is an important chapter of 
nuclear astrophysics. Specifically, the rare and fragile light nuclei 
Lithium, Beryllium and Boron (LiBeB) are not generated in the 
normal 
course of stellar nucleosynthesis (except \li7) and are, in fact, 
destroyed in stellar interiors. This characteristic is reflected in the 
low 
abundance of these simple species. Up to recently, the most 
plausible 
interpretation was that Galactic Cosmic Rays (GCR) interact with 
interstellar CNO to form LiBeB. Other origins have been also 
identified: primordial and stellar (\li7) and  supernova 
neutrino 
spallation (\li7 and \b11). In contrast, \be9, \b10
and \li6 are pure spallative products. This last isotope 
presents a 
special interest since the \li6/\li7 ratio has been 
measured 
recently in a few halo stars offering a new constraint on the early 
galactic evolution of light elements.
Optical measurements  of the beryllium and boron abundances 
in halo stars have been achieved by the 10 meter KECK telescope 
and 
the Hubble Space Telescope. These observations indicate a quasi 
linear 
correlation between Be and B vs Fe, at least at low metallicity, 
which, 
at first sight, is contradictory to a dominating GCR 
 origin of the light elements which  predicts a 
quadratic 
relationship.
As a consequence, the theory of the origin and  evolution of 
LiBeB nuclei has to be refined. 
Aside GCRs, which are accelerated in the general interstellar 
medium (ISM) and create LiBeB through the break up of CNO by 
fast 
protons and alphas, Wolf-Rayet stars (WR) and core collapse 
supernovae (SNII) grouped in superbubbles could produce copious 
amounts of light elements via the fragmentation in flight of rapid 
carbon and oxygen nuclei colliding with H and He in the ISM. In 
this 
case, LiBeB would be produced independently of the interstellar medium 
chemical composition and thus a primary origin is expected.
These different processes are discussed in the framework of a 
galactic evolutionary model. More spectroscopic observations 
(specifically of O, Fe, Li, Be, B) in halo stars are required for a better 
understanding of the relative contribution of the various mechanisms. 
Future tests on the injection and acceleration of nuclei by supernovae and 
Wolf Rayet relying on gamma-ray line astronomy will be 
invoked in 
the perspective of  the European INTEGRAL satellite.

\end{abstract}

\section{Introduction}

A general trend in nature is that complex nuclei are not proliferating: the 
abundance of the elements versus the mass number draws a 
globally 
decreasing curve. In the whole nuclear realm, LiBeB
 are exceptional since they are both simple and rare. 
Typically, in the Solar System, Li/H = 2. $10^{-9}$, B/H = 7. $10^{-10}$,
 Be/H = 2.5 $10^{-11}$ (Anders and Grevesse 1989). Indeed,
 they are rare because 
they 
are fragile and apparently a selection principle at the nuclear level 
has 
operated in nature. Due to the fact that 
nuclei with mass  5 and 8  are unstable, the Big-Bang  
nucleosynthesis (BBN)
has stopped at A = 7, and primordial thermonuclear fusion has been unable to 
proceed beyond lithium. The standard BBN 
is hoplessly ineffective in generating \li6, \be9, \b10, \b11 (fig 1).
 Thus, stars were necessary to pursue the nuclear evolution 
 bridging the gap between \he4 and \c12 much later, through 
nuclear fusion.

\begin{figure*}[htbp]
\vspace{5.5cm}
\includegraphics{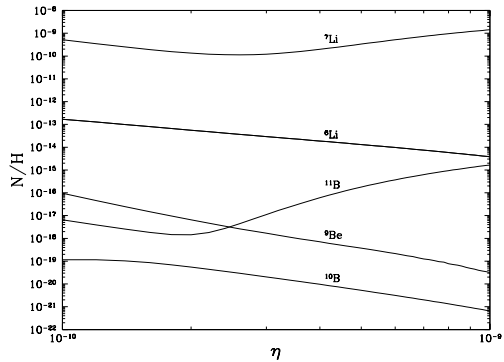}
\caption{Big Bang Nucleosynthesis of Lithium, Beryllium and Boron vs photon
 over baryon ratio}
\label{fig-paris_1}
\end{figure*}

  Stellar nucleosynthesis, quiescent or explosive, 
forge the whole variety of nuclei from C to U but they destroy 
LiBeB in the interior of stars, except \li7 which is produced in AGB and novae.
 The destruction temperature are 2, 2.5, 3.5 and 
 5.3 millions of degrees for \li6, \li7, \be9, and \b10 respectively.
Finally,  \li7 
 and \b11 could be produced by neutrino spallation in carbon shells of core
 collapse supernovae (Woosley \etal 1990, Vangioni-Flam \etal 1996); however,
 this mechanism is particularly uncertain depending strongly on the neutrino
 energy distribution.
 		
It is clear that another source is necessary to 
generate at least \li6, \be9, \b10 and this non thermal mechanism is
 the break up of heavier species 
(CNO, mainly) by energetic collisions, also called spallation.
  
The LiBeB story has been rich and moving. The genesis of LiBeB 
was so 
obscure to Burbidge \etal (1957) that they called X the process 
leading to 
their production. Then came Hubert Reeves and his students. In a 
seminal work, Meneguzzi, Audouze and Reeves (1971) identified the 
production process, i.e. the Galactic Cosmic Rays - Interstellar 
Medium interaction. Exploiting the  fast p,$\alpha$ in the GCRs interacting
 with CNO in the ISM, they were able to make quantitative estimates of the 
LiBeB 
production on the basis of cross section 
measurements notably made in Orsay (Raisbeck and Yiou, 1971, 1975).
 However, this estimate, based on the local and present observations (LiBeB 
 and CNO abundances, cosmic ray flux and spectrum) was based on an extrapolation
 over the whole galactic lifetime assuming that all the parameters are
 constant. This result accounted fairly well for the cumulated light
 element abundances
 but obviously not for their evolution which, at that time, was unknown.
The pertinence  of their idea is illuminated by the simple and
 beautiful fact that the hierarchy of the abundances \b11 $>$ \b10 $>$ \be9 
is 
reflected in the cross sections (Read and Viola, 1984). This is another 
proof that nature follows the rules of nuclear physics.
\li6, \be9 and \b10 were nicely explained but problems were 
encountered 
with \li7 and \b11. The calculated \li7/\li6 ratio was 1.2 against 12.5 
in 
meteorites. Stellar sources of \li7 appeared necessary. The estimated 
\b11/\b10 ratio was 2.5  instead of 4 in meteorites. An ad-hoc hypothesis 
drawing on unobservable low energy proton  operating through
 the \n14(p,x)\b11 reaction was advocated (see Reeves 1994 for a review).

New measurements of Be/H and B/H from KECK and HST, together 
with [Fe/H] (Rebolo \etal 1988, Gilmore \etal 1992, Duncan \etal 1992,
 Boesgaard and King 1993, Ryan \etal 1994, Duncan \etal 1997, Garc\'ia-L\'opez
 \etal 1998) in very low metallicity halo stars came to set 
strong 
constraints on the origin and evolution of  light isotopes.

 The evolution of BeB was suddenly known  over about 10 Gyr, taking [Fe/H] 
as 
an evolutionary index.
A compilation of Be and B data is presented in fig 2. The most 
striking 
point is that log(Be/H) and log(B/H) are both quasi proportional to [Fe/H],
 at least up 
to  [Fe/H] = -1 and that the B/Be ratio lies in the range 10 - 30
 (Duncan \etal 1997).

\begin{figure*}[htbp]
\vspace{5.5cm}
\includegraphics{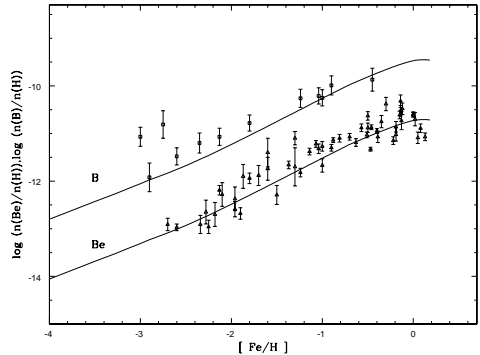}
\caption{Beryllium and Boron evolution vs [Fe/H]}
\label{fig-paris_2}
\end{figure*}

This linearity came as a surprise since a quadratic 
relation was expected from the GCR mechanism. It was a strong
 indication that the standard GCRs are not the main producers of LiBeB in
 the early 
Galaxy. A new mechanism of primary nature was required to
 reproduce these observations: low energy fast CO nuclei
 produced and accelerated by massive stars (WR and SN II) fragment on H and He
 at rest in the ISM. This low energy component (LEC) has the advantage of
 coproducing Be and B in good agreement with the ratio observed in Pop II
 stars, (figure 2  and Vangioni-Flam \etal 1998a). 

 A primary origin , in this 
language, means a production rate independent of the interstellar 
metallicity. In this case, the cumulated abundance of a given light 
isotope L is approximately 
 proportional to Z. At variance, standard GCRs offer a 
secondary mechanism because it should depend both on the CNO abundance of the
 ISM at a given time and on the intensity of cosmic ray flux, itself assumed
 to be proportional to the SN II rate.

 Note however, the two discrepant points in the boron diagram at the lowest 
 [Fe/H]. This is mainly due to the huge NLTE correction on the data (Kiselman
 and Carlsson 1996) that increases
 the departure from a straight line. It is important to take a carefull
 look to this delicate correction.  

 The Be-Fe and B-Fe correlations taken at face value show
 a contradiction between 
theory and observation.
But, since oxygen is the main progenitor of BeB, the
 apparent linear relation between BeB and Fe could be misleading if O were
 not strictly proportional to Fe (Israelian \etal 1998 and Boesgaard \etal
 1998). Thus the pure primary origin of BeB could be questionned
 (Fields and Olive
 1998). However, the oxygen measurements themselves are confronted to many
 difficulties (Mac Williams 1997, Cayrel, Spite and Spite, private
 communication). On the theoretical side, the situation is not better. The 
 [$\alpha$/Fe] vs [Fe/H] where $\alpha$ = Mg, Si, Ca, S, Ti (Cayrel 1996, Ryan
 \etal 1996) show a plateau from about [Fe/H] = -4 to -1.  On nucleosynthetic
 grounds, it would be surprising that oxygen does not follow the Si and 
 Ca trends.
 Moreover, using the published nucleosynthetic yields (Woosley and Weaver 1995, 
 Thielemann, Nomoto and Hashimoto 1996) it is impossible to fit the log(O/H) vs
 [Fe/H] relation of Israelian \etal (1998) and Boesgaard \etal (1998) since
 the required oxygen yields are unrealistic. Thus the subject is controversial.

Concerning lithium, a  compilation of the data is
 shown in Lemoine \etal (1997) and 
Molaro et 
al (1997). The Spite plateau extends up to [Fe/H] = -1.3. Beyond, Li/H is 
strongly increasing until its solar value of 2. $10^{-9}$.

 A stringent 
constraint to any theory of Li evolution is avoiding to cross the Spite's 
plateau 
below [Fe/H] = -1. Accordingly, the Li/Be production ratio should 
be 
less than about 100. 

Recent measurements of \li6 have been made successfully in two halo 
stars, HD84937 and BD +26 3578 at about  [Fe/H] =  -2.3 ( Hobbs and 
Thorburn 
1997, Cayrel \etal 1998, Smith \etal 1998), yielding
\li6/\li7 about  0.05.
The great interest of \li6, besides of being an indicator of stellar 
destruction (Pinsonneault \etal 1998, Chaboyer 1998, Cayrel \etal 
1999, 
Vauclair and Charbonnel 1998) is to represent a pure spallation
 product as \be9.

 Lithium-6 has different sources, a secondary one and two primary ones: i)
 fast (p,$\alpha$) on CNO at rest (this secondary 
process related to GCRs should not be efficient in the early galaxy),
 ii) fast CO on H, He (primary) and  iii) specifically $\alpha$ + $\alpha$
 at low energy (primary). 
 The non thermal fusion reaction $\alpha$ + $\alpha$  produces almost equal 
amounts 
of \li6 and \li7 at low energy and the cross section above 100 
MeV is specially low (Read and Viola 1984). The second and third processes are
 associated to LEC. Consequently, this  LEC is specially fertile in
 lithium isotopes.

Preliminary estimates of the \li6/\be9 ratio have been
 performed for few Pop II stars but with a large uncertainty (20 - 80). This 
 range of values is much higher than the \li6/\be9 ratio generated by the 
present GCR (6). This could indicate that this ratio is varying all along the
 galactic evolution (Vangioni-Flam \etal 1998b, Fields and Olive 1998).

 To summarize, we can give  six 
 observational constraints on LiBeB evolution :

                 1. Be and B proportional to Fe
 
                 2. Li/Be $<$  100 up to [Fe/H] = -1

                 3. B/Be = 10-30
 
                 4. \b11/\b10 = 4 at solar birth

                 5. \li7/\li6 = 12.5 at solar birth

                 6. \li6/\li7 = 0.05  and \li6/\be9 = 20 to 80 (to be
 confirmed) at [Fe/H] about -2.3

We recall that the observational O - Fe relation is central to the
 interpretation since specifically the production of Be is related to O.
 Most of the observers find oxygen proportional to Fe.

\section{Basic physical parameters of non thermal nucleosynthesis of 
LiBeB}

Four parameters are influential to the spallative production of light elements: 
the 
reaction cross sections, the energy spectrum of fast nuclei, the 
composition of the beam and the composition of the target. 

 Cross sections are well measured (Read and Viola 1984)
 and have been 
updated recently by Ramaty \etal (1997).

 The adopted spectra are  of two kinds:

1.  GCR: N(E)dE  = k$E^{-2.7}$ above a few GeV/n  with a 
flatenning below (e.g. Lemoine \etal 1998). 

2. LEC : Shock wave acceleration with a cut at Eo of the form 
N(E) dE = k$E^{-1.5}$ exp(-E/Eo) dE (Ramaty \etal 1996), propagated in the ISM.

 The source composition of GCR is well determined (e.g. Du Vernois 1996).
 It is p and $\alpha$ rich (H/O = 200, He/O = 20)
 contrary to the possible  source composition of the LEC.
 The most obvious contributors to LEC are supernovae, 
Wolf-Rayet and mass loosing stars (Cass\'e \etal 1995, Ramaty \etal
1996, 1997, 1998). It is worth noting that in the  early galaxy,
supernovae
play a leading role since at very low metallicity the stellar winds are 
unsignificant.
Table 1 shows a sample of compositions  used by different authors 
: solar system (SS) for comparison (Ramaty \etal 1996 from
 Anders and Grevesse 1989), cosmic ray source (CRS) (Ramaty \etal 1996 from
 Mewaldt 1983), wind of massive stars (W40) (Parizot \etal 1997 from Meynet
 \etal 1994), composition of grain products (GR) (Ramaty \etal 1996 and 
 Lingenfelter \etal 1998), 40 Mo
 supernova at Z = $10^{-4}$ Zo (Parizot \etal 1999 from Woosley and Weaver 1995)
, 35 Mo supernova of solar metallicity (Ramaty \etal 1996 from Weaver and
 Woosley 1993). The two supernovae, though at different metallicities,
 (SN40 and SN35),  give similar yields due to the fact
 that metallicity dependent mass loss has not been taken into account. 
 Resulting elemental and isotopic ratios (B/Be, \b11/\b10, \li6/\be9)
 for different compositions and Eo can be found in Ramaty \etal (1996) and
 Vangioni-Flam \etal (1997). 
 
\begin{table}[ht]
\centerline {\sc{\underline {Table 1 : Source Composition}}}
\vspace {0.1in}
\begin{center}
\begin{tabular}{|lcccccc|}        \hline \hline
 Element & SS	 & CRS & W40  & GR & SN40(low Z)  & SN35(Zo)  \\ \hline
H & 1200 & 220 & 80 & 2 & 37 & 27 \\
He & 120 & 22 & 25 & 0 & 8.8 & 7.6 \\
C & 0.47 & 0.87 & 1.6 & 0.3 & 0.09 & 0.08 \\
N & 0.13 & 0.04 & - & 0.03 & - & -  \\
O & 1 & 1 & 1 & 1 & 1 & 1 \\
\hline
\end{tabular}
\end{center}
\end{table}

Note that N is unsignificant and that  Type II supernovae are 
O-rich whereas the winds of massive stars are C and He-rich. These 
abundance differences are important since the  highest \b11/\b10 
ratios 
are  produced by C-rich beams through \c12(p,x)\b11 and the highest \li6/\be9 
 ratios are produced by He and O rich compositions.

The fourth parameter, i.e., the composition of the target (ISM) varies from
 the birth of the galaxy up to now.
 The extensive study of 
Ramaty \etal (1996) and Vangioni-Flam \etal (1997) shows that there are only
 slight differences in the results
 when the ISM metallicity is varied between 0 (early galaxy) and Zo (now), 
except perhaps concerning the \b11/\b10 ratio.

\section{LiBeB production mechanisms and galactic evolution}

Analyzing  all the physical parameters discussed above, two main 
 LiBeB producers emerge, the first one is the standard GCR (spectrum 1) in which
 fast p,$\alpha$ nuclei interact with CNO in the ISM. This process seems unable
to produce sufficient amounts of LiBeB at the level observed in the halo 
stars (Vangioni-Flam \etal 1996). A recent study (Fields and Olive 1998) based
 on the O - Fe relation derived by Israelian \etal (1998) and Boesgaard \etal
 (1998) at low metallicity (still controversial) try to fit the observational
 constraints with
 a pure standard GCR (secondary production) component, but has problems with
 the B/Be ratio among other things.

The second one, reverse to the GCR mechanism, invokes fragmentation of CO nuclei
 in flight by collision with H and He in the ISM.
 Massive stars  are able in principle 
 to furnish freshly synthesized C and O and accelerate them 
via the shock waves they induce. This mechanism is related to superbubbles (S)
 through a scenario proposed by Bykov (1995), Parizot (1998) and 
Vangioni-Flam \etal (1998a). Here only the most massive stars ( greater than
 about 60 Mo) 
contribute due their short lifetime.
Originally proposed in relation with the observation of gamma ray 
line 
from Orion it remains as a distinctive possibility after the 
withdrawal 
of the COMPTEL results on this molecular complex (Bloemen \etal 1998) and the 
announcement of a possible excess of gamma rays in the 3-7 MeV 
range from the Vela region. Possible observations in X-rays seem to 
 substantiate this scenario (Tatischeff \etal 1998).
 The observation or non observation of C, O 
lines at 4.4 and 6.1 MeV and of the Li-Be feature close to 500 keV by the 
INTEGRAL satellite (Winkler 1997) will be the  strongest test of the 
superbubble hypothesis (Parizot \etal 1997).
  
The acceleration of grain debris in supernovae (scenario G) is another 
version of a primary mechanism (Lingenfelter \etal 1998, Ramaty,
 these proceedings). The G model leads 
to  quite different predictions concerning the evolution of Be, B  and 
\li6 
only at very low metallicity (Vangioni-Flam \etal 1998a,b) because all SN II
 (10 - 100 Mo) are implied rather than the most massive ones in the S scenario.

Finally, neutrino spallation (Woosley \etal 1990, Woosley and Weaver 1995,
 Vangioni-Flam \etal 
1996) is helpful to increase the \b11/\b10 ratio up to the value 
observed in meteorites. 
It is also a primary process since it implies the break up of \c12 
within 
supernovae and not in the ISM. However it cannot be the unique 
mechanism to produce light elements since it does not produce \be9. Moreover,
 it has been shown that its contribution is only marginal (Vangioni-Flam \etal 
 1996).

These different mechanisms are included in a galactic 
evolutionary model (Vangioni-Flam \etal 1996, 1998a,b) to follow the whole
 evolution of each isotope.

 Concerning beryllium and boron, in this context,   
the main results are the following: the quasi-linearity (Be-B vs Fe) is
easily  reproduced (fig 2).
Standard GCR contribute no more than about 30 per cent to Solar System values. 
The B/Be ratio is in the range 10-30 as observed. The value 30 leaves enough 
room for neutrino spallation to reach \b11/\b10  = 4 at solar birth.

The \li6/H ratio can be explained in the 
framework 
of the same superbubble model (Vangioni-Flam \etal 1998b, Cayrel \etal 
1998) this
without piercing the Spite plateau (fig 3). In this figure, showing
 the evolution of
 \li6/H vs [Fe/H], it can be seen that GCR is overwhelmed by LEC.
 The decrease of the  \li6/\be9
 ratio could be  explained in terms of the variation of the composition of 
superbubbles in the course of the galactic evolution, being O rich at 
start 
due to SNII and becoming more and more C rich due to the 
increasing contribution of mass loosing stars (Vangioni Flam \etal 
1998b). Morover, the evolutionary curve of \li6 crosses the halo observations
 (fig 3)  meaning that \li6 is almost essentially intact in the envelope
 of stars in
 which it is measured. \li7 in turn, more tightly bound than \li6,
 is even less destroyed,
 thus the mean value of the Spite plateau reflects nicely the Big Bang \li7
 abundance. This reinforce the use of \li7 as a cosmological baryometer.

\begin{figure*}[htbp]
\vspace{5.5cm}
\includegraphics{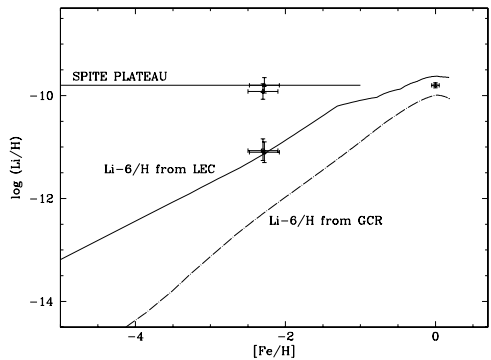}
\caption{Evolution of  Lithium vs [Fe/H]}
\label{fig-paris_3}
\end{figure*}

\section{ CONCLUSION}

 Recent LiBeB observations indicate that a primary component is probably at 
work in the early Galaxy, presumably related to core collapse  
supernovae. Promising scenarios have been presented implying respectively 
the 
acceleration of freshly synthesized C and O in superbubbles and/or 
grains debris around supernovae (see also Ramaty in these proceedings).

 New \li6 observations put strong constraints on the  
composition 
and  spectrum of an early population of fast particles. The superbubble model
 is also able to reproduce the \li6 observation, until the local meteoritic 
 value (fig 3). 
A low energy component originally  O rich and becoming progressively
 C rich due to the strengthening of stellar winds at increasing
 metallicities is required
to explain the high \li6/\be9 observed in a  few halo stars with respect to 
the one measured in meteorites. But a definitive
 conclusion should wait confirmation.
 
Essentially no destruction of \li6 and \li7 is implied by the evolutionary
 curve of \li6. As a consequence, \li7 is a good baryonic density indicator
 for cosmology.  
The needs for the future are the following:
 
 On the theoretical side it would be necessary:

 i) to check 
NLTE corrections on B abundances since two bothering points remain 
at very low Z.

 ii) to develop and refine SN II models, specially at very low Z and 
high mass, M>60 Mo.

 On the observational side, it would be desirable to get 
measurements 
of \li6, \li7, \be9, B, O, Fe in the same halo stars and to get \b11/\b10 
ratios in various stars. A first step in this goal has been 
accomplished by 
Rebull \etal (1998). 

Finally, the 
observation of C,O and Li-Be gamma-ray lines are important 
objectives of the INTEGRAL satellite which will open up an 
European era in gamma-ray astronomy (Parizot \etal 1997,
 Cass\'e \etal 1998).

\section{Aknowledgements}
We have been very much honored to participate to the celebration 
of 
the scientific work of  Giusa and Roger Cayrel, who are an example for 
every researcher in Astronomy. Moreover, we thank warmly 
Monique and Francois Spite for the excellent organization of the 
meeting. This work was supported in part by the PICS 319, CNRS.

\end{document}